\documentclass[prl,amsfonts,amssymb,floats,twocolumn,superscriptaddress,aps]{revtex4}

\usepackage[final,dvips]{epsfig}

\begin{document}

\title[]
{Numerical Renormalization Group for Bosonic Systems \\
and Application to the
Subohmic Spin-Boson Model
}

\author{Ralf Bulla}
\affiliation{\mbox{Theoretische Physik III, Elektronische Korrelationen und
Magnetismus, Universit\"at Augsburg, Germany}}
\author{Ning-Hua Tong}
\affiliation{\mbox{Theoretische Physik III, Elektronische Korrelationen und
Magnetismus, Universit\"at Augsburg, Germany}}
\author{Matthias Vojta}
\affiliation{\mbox{Institut f\"ur Theorie der Kondensierten Materie,
Universit\"at Karlsruhe, Germany}}
\date{Sep 4, 2003}

\begin{abstract}
We describe the generalization of Wilson's Numerical Renormalization Group method
to quantum impurity models with a bosonic bath,
providing a general non-perturbative approach to bosonic impurity models
which can access exponentially small energies and temperatures.
As an application, we consider the spin-boson model,
describing a two-level system coupled to a bosonic bath with power-law spectral
density, $J(\omega)\propto\omega^s$.
We find clear evidence for a line of continuous quantum phase
transitions for subohmic bath exponents $0<s<1$;
the line terminates in the well-known Kosterlitz-Thouless transition at $s=1$.
Contact is made with results from perturbative renormalization group,
and various other applications are outlined.
\end{abstract}


\maketitle

The Numerical Renormalization Group method (NRG) developed by
Wilson \cite{Wil75} is a powerful tool for the investigation
of the Kondo model and its generalizations
\cite{Wil75,Kri80,Hewson,RBAdvPhys}.
In these models, a (possibly complex) impurity, such as a localized spin,
couples to a fermionic bath.
% , characterized by a density of states $\rho(\varepsilon)$.
In the case of a spin-$\frac{1}{2}$ impurity coupled
antiferromagnetically to a metallic bath, the impurity spin
is screened below a characteristic scale $T_{\rm K}$, the
Kondo temperature \cite{Hewson}.
The strength of the NRG lies in its non-perturbative nature
and the ability to resolve arbitrarily small energies \cite{Wil75}.
A variety of thermodynamic and dynamic quantities can be calculated for
a large number of impurity models in the whole
parameter space \cite{RBAdvPhys,Costi99a}.

There is, however, a very important class of models for which the NRG
method has not yet been developed: models with a coupling
of the impurity to a {\em bosonic} bath \cite{footnote}.
The intensively studied spin-boson model \cite{Leggett, Weiss}
belongs to this class; its Hamiltonian is given by
\begin{equation}
H=-\frac{\Delta}{2}\sigma_{x}+\frac{\epsilon}{2}\sigma_{z}+
\sum_{i} \omega_{i}
     a_{i}^{\dagger} a_{i}
+\frac{\sigma_{z}}{2} \sum_{i}
    \lambda_{i}( a_{i} + a_{i}^{\dagger} ) \ .
\label{eq:sbm}
\end{equation}
Here the Pauli-matrices $\sigma_j$ describe a spin, i.e., a generic two-level system,
which is linearly coupled to a bath of harmonic oscillators,
with creation (annihilation) operators $a_i^\dagger$ ($a_i$).
The bare tunneling amplitude between the two spin states $|\uparrow\rangle$
and $|\downarrow\rangle$ is given by  $\Delta$, and $\epsilon$ is an additional bias.
The $\omega_i$ are the oscillator frequencies and $\lambda_i$ the coupling strengths
between the oscillators and the local spin.
The coupling between spin and bosonic bath is completely specified
by the bath spectral function
\begin{equation}
    J\left( \omega \right)=\pi \sum_{i}
\lambda_{i}^{2} \delta\left( \omega -\omega_{i} \right) \,.
\end{equation}
Of particular interest are power-law spectra
\begin{equation}
  J(\omega) = 2\pi\, \alpha\, \omega_c^{1-s} \, \omega^s\,,~ 0<\omega<\omega_c\,,\ \ \ s>-1
\label{power}
\end{equation}
where the dimensionless parameter $\alpha$ characterizes the
dissipation strength, and $\omega_c$ is a cutoff energy.
The value $s=1$ corresponds to the case of ohmic dissipation.

The spin-boson model is a generic model describing
quantum dissipation; it has been discussed in the
context of a great variety of physical problems \cite{Leggett, Weiss}
ranging from the effect of friction on the electron
transfer in biomolecules \cite{Garg} to the description
of the quantum entanglement between a qubit and its
environment \cite{Cos03,dima,wilhelm}.
% (for further applications see Refs. \onlinecite{Leggett, Weiss}).

Considering the wealth of applications, the question
arises whether Wilson's NRG method can be exploited for this
class of models; and it is the purpose of this paper to show that this
is indeed the case.
What we have in mind here is the direct mapping of models like (\ref{eq:sbm})
to a semi-infinite chain form typical for the NRG
\cite{Wil75}.
As described below, bosonic operators constitute the
sites of the chain, and the Hamiltonian is solved by iterative
numerical diagonalization \cite{footalex}.
This approach is different from previous NRG calculations
in Refs.~\onlinecite{Costi,Cos03}, where
the mapping of the spin-boson model (\ref{eq:sbm}) to
the anisotropic {\em fermionic} Kondo
model was employed -- such a mapping is restricted to the
ohmic case $J(\omega)\propto\omega$.
% Nevertheless, the NRG-calculations of \cite{Costi} give the most precise information
% for dynamic quantities of the ohmic SBM, and we will refer to
% these further below (do we?).

{\em Bosonic NRG. }
Let us now describe the generalization of the NRG
method to a bosonic bath with a continuous spectrum.
The strategy is similar to the
one used for the Kondo or single-impurity Anderson model
\cite{Wil75,Kri80}.
There are, however, important differences which we outline
here; a more detailed discussion % of the technical steps
will appear elsewhere.
Here we present explicit equations for the spin-boson model (\ref{eq:sbm});
the generalization to other impurity models or multiple bosonic baths
is straightforward.

% A convenient starting point for the NRG is
We start from the following form of the model (\ref{eq:sbm}):
\begin{equation}
 H= H_{\rm loc} +
  \int\limits_{0}^{1} {\rm d}\varepsilon \, g(\varepsilon) a_{\varepsilon}^{\dagger} a_{\varepsilon}
 + \frac{\sigma_{z}}{2} \int\limits_{0}^{1} {\rm d}\varepsilon \, h(\varepsilon )
   (a_{\varepsilon} + a_{\varepsilon}^{\dagger} )
\end{equation}
with $H_{\rm loc} = -\Delta\sigma_{x}/2 + \epsilon\sigma_{z} / 2$.
In this model, $g(\varepsilon)$ characterizes the dispersion
of a bosonic bath in a one-dimensional representation,
with upper cutoff 1 for $\varepsilon$. The coupling
between the spin and the bosonic bath is given by $h(\varepsilon)$.
These two energy-dependent functions are related to the spectral
function $J(\omega)$ via
\begin{equation}
  \frac{1}{\pi}J(x)=\frac{d \varepsilon(x)}{d x} h^{2} \left[ \varepsilon(x) \right]
  \quad \left( x \in \left[0, \omega_{c} \right] \right) \, ,
\label{eq:h}
\end{equation}
where $\varepsilon(x)$ is the inverse function of $g(x)$, $g[\varepsilon(x)]=x$.
As discussed in Ref.~\onlinecite{BPH} % in the context of
for the Anderson model,
Eq.~(\ref{eq:h}) does not uniquely determine $g(x)$ and $h(x)$,
and a specific choice for $h(x)$ is used to simplify the
calculations.

The NRG procedure starts by dividing the energy interval
$[0,1]$ into intervals $[\Lambda^{-(n+1)},\Lambda^{-n}]$
($n=0,1,2,\ldots$). An orthonormal set of functions
$\psi_{np}(\varepsilon) \propto e^{i\omega_n p \varepsilon}$ is
introduced for each interval
so that the operators $a_\varepsilon$ can be represented in
this basis. Choosing $h(\varepsilon)$ as constant in each
interval \cite{BPH} and dropping the $(p\!\neq\!0)$-components as in
\cite{Wil75,Kri80}, the Hamiltonian of the spin-boson model then takes the
following form:
\begin{eqnarray}
H &=& H_{\rm loc} + \sum\limits_{n=0}^{\infty} \xi_{n}a_{n}^{\dagger}a_{n}
%\nonumber\\
  + \frac{\sigma_z}{2\sqrt{\pi}}
  \sum\limits_{n=0}^{\infty} \gamma_{n} \left(a_{n}+a_{n}^{\dagger} \right),
\nonumber \\
    \xi_{n}\!&=& \! \gamma_n^{-2}
    \int_{\Lambda^{-(n+1)}\omega_c}^{\Lambda^{-n}\omega_c}
    \!\!\!\!\!\!\!\!\!\!{\rm d}x \,J(x)x \, ,~~
%   \end{equation}
%and
%   \begin{equation}
    \gamma_{n}^2=
    \int_{\Lambda^{-(n+1)}\omega_c}^{\Lambda^{-n}\omega_c}
    \!\!\!\!\!\!\!\!\!\!{\rm d}x \, J(x) .
\end{eqnarray}
The transformation to a semi-infinite chain form yields
\begin{eqnarray}
  H_{\rm sem}&=& H_{\rm loc} +
   \sqrt{\frac{\eta_0}{\pi}} \frac{\sigma_{z}}{2} \left(b_{0}+b_{0}^{\dagger} \right) \nonumber\\
   &+&\sum\limits_{n=0}^{+\infty} \left[ \epsilon_{n}b_{n}^{\dagger}b_{n}
         +t_{n}\left( b_{n}^{\dagger} b_{n+1}+b_{n+1}^{\dagger} b_{n}\right)
      \right]  \label{eq:Hsem}
\end{eqnarray}
with $\eta_0 = \int{\rm d}x\,J(x)$.
The spin now couples to the first site of the bosonic chain
only, and the remaining part of the chain is characterized
by on-site energies $\epsilon_n$ and hopping parameters $t_n$.
% in analogy to the fermionic NRG.
The parameters $\eta_0$, $\epsilon_n$ and $t_n$ can be calculated numerically
from a given spectral function $J(\omega)$ \cite{BPH}.
Note that here the spectrum is restricted to positive
frequencies; this results in hopping matrix elements falling
off as $t_n\propto \Lambda^{-n}$
(which allows to work with $\Lambda=2$ keeping a relatively small number of states),
in contrast to the fermionic case where
the discretization is performed for both negative and
positive energies, in this case $t_n\propto \Lambda^{-n/2}$.
The on-site energies also fall off as $\epsilon_n\propto \Lambda^{-n}$
so that a fixed but $s$-dependent ratio $t_n/\epsilon_n$ emerges for large $n$,
where $s$ is the bath exponent in (\ref{power}).

The Hamiltonian (\ref{eq:Hsem}) is solved by iterative numerical
diagonalization % , in close analogy to the case of a fermionic bath
\cite{Wil75,Kri80}.
At each step, one bosonic site of the chain is added.
The infinite bosonic Hilbert space has to be cut off,
by restricting the basis of each new bosonic site to a
finite number of states $N_b$.
After diagonalizing the enhanced cluster, the $N_s$ lowest lying
many-particle states are kept, and the procedure is repeated
\cite{states}.
The calculation of static and dynamic observables can be done in analogy
to the fermionic NRG.
In general, and as known from the fermionic case, the accuracy of the
cutoff procedure has to be tested for each application,
and we will show results below.

%%%%%%%%%%%%%%%%%%%%%%%%%%%%%%%%%%%%%%%%%%%%%%%%%%%%%%%%%%%%%%%%%%%%%%%%%

\begin{figure}[!t]
\epsfxsize=3.5in
\centerline{\epsffile{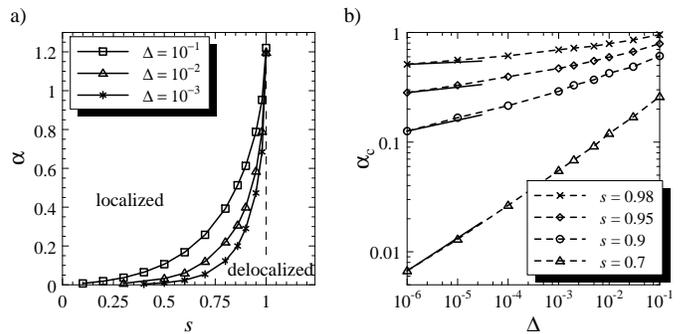}}
\caption{
a) Phase diagram for the transition between delocalized
($\alpha < \alpha_c$) and localized phases ($\alpha > \alpha_c$)
of the spin-boson model (\protect\ref{eq:sbm})
for bias $\epsilon=0$ and various values of $\Delta$,
deduced from the NRG flow.
b) $\Delta$-dependence of the critical coupling $\alpha_c$ for various
values of the bath exponent $s$. The dashed lines are guides to the
eye, the solid lines are fits to Eq.~(\protect\ref{alpcrg})
using the $\Delta<10^{-7}$ points only.
For $s$ close to 1 the asymptotic regime is only reached for
very small $\Delta$.
NRG parameters here and in Figs.~\protect\ref{fig:flow}--\protect\ref{fig:exp}
are $\Lambda\!=\!2$, $N_b\!=\!8$, and $N_s\!=\!100$.
}
\label{fig:phd}
\end{figure}

{\em Application to the spin-boson model. }
To investigate the feasibility of the bosonic NRG, we performed
extensive calculations for the spin-boson model with
bath exponents $0<s\leq 1$, bias $\epsilon\!=\!0$, and $\omega_c\!=\!1$.
In the ohmic case $s\!=\!1$, it is known that a Kosterlitz-Thouless
quantum transition separates a localized phase at $\alpha \geq \alpha_c$
from a delocalized phase at $\alpha<\alpha_c$ \cite{Leggett,Weiss}.
In the localized regime, the tunnel splitting between the two
levels renormalizes to zero, whereas it is finite in the delocalized
phase.
For $\Delta \ll \omega_c$ the transition occurs at $\alpha_c=1$.

The subohmic case \cite{dima,KM} is less clear.
For $\Delta/\omega_c\to 0$ the system is localized for any non-zero
coupling, however, the behavior at finite $\Delta$ was not discussed
in Refs.~\onlinecite{Leggett,Weiss}.
For large $\Delta$ a delocalized phase was argued to exist \cite{spohn,KM},
and Ref.~\onlinecite{KM} proposed a first-order transition scenario.
In the following, we shall resolve this issue and show that a
continuous transition with associated critical behavior occurs
for all $0<s<1$.

Notably, the spin-boson model can be mapped onto a one-dimensional
Ising model with long-range couplings falling off as $r^{-s-1}$;
the localized phase of the spin-boson model then corresponds to
the ordered phase of the Ising magnet \cite{Leggett}.
As shown by Dyson \cite{dyson}, this Ising model features a transition
for $0<s\leq 1$, but the results for $s<1$ have not been systematically
carried over to the spin-boson model so far.

Our NRG calculations provide clear evidence for a phase transition
in the spin-boson model for all $0<s\leq 1$, which is continuous
for $0<s<1$ and of Kosterlitz-Thouless type for $s=1$.
The numerical results are summarized in Fig.~\ref{fig:phd}a, which shows the
phase boundaries determined from the NRG flow for fixed NRG parameters
$\Lambda\!=\!2$, $N_b\!=\!8$, and $N_s\!=\!100$ \cite{states}.
(No transition occurs for $s>1$: the system is always delocalized.)
As displayed in Fig.~\ref{fig:phd}b, the critical coupling $\alpha_c$
closely follows a power law as function of the bare tunnel splitting,
$\alpha_c \propto \Delta^x$ for small $\Delta$, with an $s$-dependent
exponent $x$.
Our data are consistent with $x = 1-s$, see below.

\begin{figure}[!t]
\epsfxsize=2.8in
\centerline{\epsffile{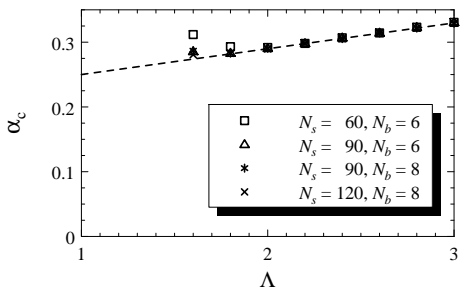}}
\caption{$\Lambda$ dependence of the critical coupling
$\alpha_c$ at $s=0.9$, $\Delta=10^{-3}$,
for various NRG parameters $N_s$ and $N_b$ \protect\cite{states}.
The dashed line is a linear fit to the $N_s=120$, $N_b=8$ data
in the range $2\leq\Lambda\leq 3$.
}
\label{fig:lambda}
\end{figure}

In the ohmic case, $s=1$, the critical $\alpha_c$ approaches a {\em finite}
value as $\Delta \to 0$.
For the quoted NRG parameters we find $\alpha_c \approx 1.18$,
being slightly larger than the established value
$\alpha_c(s=1,\Delta \to 0)=1$.
This deviation is solely due to the NRG discretization;
calculations with different $\Lambda$ show that
in the limit $\Lambda\to 1$ we recover $\alpha_c=1$.
The general behavior is illustrated in Fig.~\ref{fig:lambda},
which shows $\alpha_c$ for fixed $\Delta$ and $s=0.9$.
Keeping $\Lambda$ fixed, we observe a rapid convergence of $\alpha_c$
with increasing $N_b$ and $N_s$.
As expected from the iterative diagonalization scheme,
the values of $N_b$ and $N_s$ necessary for convergence
increase with decreasing $\Lambda$ \cite{states}.
The converged data for $\alpha_c(\Lambda)$
show a linear $\Lambda$ dependence in the range
$1.8<\Lambda<3$, with a deviation of about 15\% at
$\Lambda = 2$ from the extrapolated $\Lambda\to 1$ value.
The same holds for the ohmic case (data not shown)
and the extrapolation results in
$\alpha_c(s=1,\Delta=10^{-4},\Lambda\to 1) = 0.99 \pm 0.02$;
our data are consistent with the RG result
$\alpha_c = 1 + {\cal O}(\Delta/\omega_c)$ \cite{Leggett}.

\begin{figure}[!t]
\epsfxsize=3.6in
\centerline{\epsffile{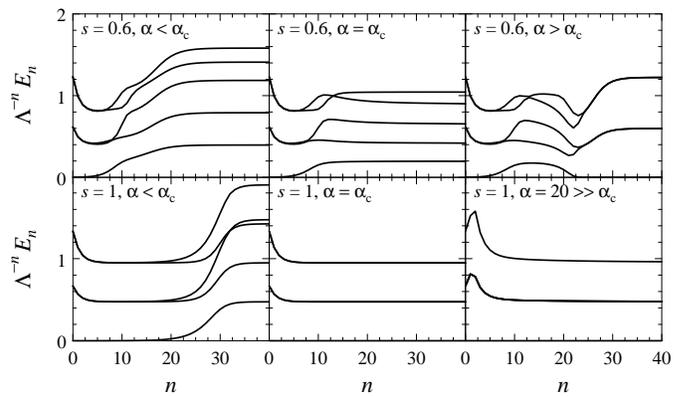}}
\caption{
Sample NRG flow diagrams for the
subohmic (upper panel, $s=0.6$) and
the ohmic case (lower panel, $s=1$),
close to the phase transition (except for the lower right panel).
}
\label{fig:flow}
\end{figure}

The NRG flow of the many-particle levels of the Hamiltonian,
displayed in Fig.~\ref{fig:flow}, can be used to analyze the
low-temperature behavior.
For all values of $0<s<1$, we can identify two stable fixed points,
corresponding to the localized and delocalized phases of the
impurity spin, and a third NRG fixed point, which is
infrared unstable and corresponds to a critical fixed point.
In contrast, for $s=1$ we find (in addition to the delocalized fixed point)
a {\em line} of fixed points for $\alpha\geq\alpha_c$,
and {\em no} critical fixed point, as
expected for a Kosterlitz-Thouless transition.

\begin{figure}[!b]
\epsfxsize=3.4in
\centerline{\epsffile{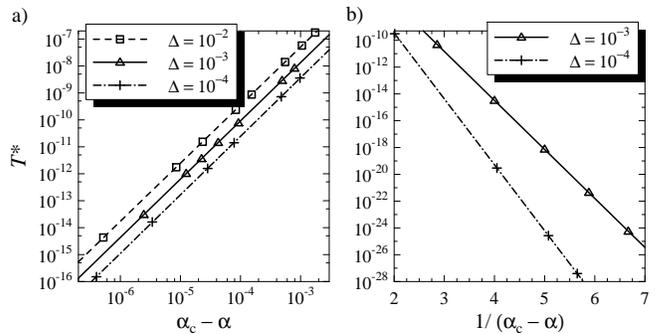}}
\caption{
Crossover scale $T^\ast$ in the vicinity
of the phase transition for $\alpha<\alpha_c$, defined here through
the first excited NRG level $\Lambda^{n} E_{n1}(T^\ast)\!=\!0.3$.
a) Subohmic case $s=0.8$, with power law fits.
b) Ohmic case $s=1$, with exponential fits.
}
\label{fig:tstar}
\end{figure}

The energy scale $T^\ast$, describing the crossover from the
critical to a stable fixed point, is shown in Fig.~\ref{fig:tstar}.
For $0<s<1$, $T^\ast$ is found to vary in a power-law fashion with
the distance from criticality,
$T^\ast \propto |\alpha-\alpha_c|^{\nu z}$,
where we have introduced the correlation length and dynamical
exponents $\nu$ and $z$ -- note that
$1/(\nu z)$ is nothing but the scaling dimension of the leading relevant operator
at the critical fixed point.
Fig.~\ref{fig:tstar}a nicely shows that $\nu z$ is independent of $\Delta$
for fixed $s$, further supporting the existence of a continuous quantum phase
transition with universal behavior.
In the ohmic case $s=1$, $T^\ast$ varies exponentially with the distance from
the critical coupling, $\ln T^\ast \propto 1/(\alpha_c-\alpha)$,
as expected (Fig.~\ref{fig:tstar}b).

\begin{figure}[!t]
\epsfxsize=3.5in
\centerline{\epsffile{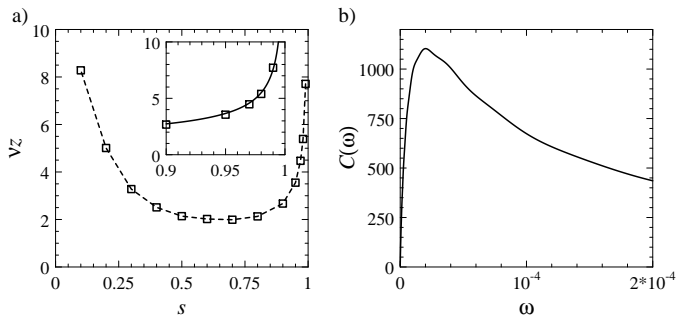}}
\caption{
a) NRG results for the critical exponent $\nu z$, characterizing
the vanishing of the crossover energy $T^\ast$ near the critical point,
as function of the bath exponent $s$.
Inset: data near $s=1$ together with the
function $1/\sqrt{2(1-s)} + 0.5$, Eq. (\protect\ref{nuzrg}).
b) Spin autocorrelation function $C(\omega)$ for $s=0.6$, $\Delta=10^{-2}$,
and $\alpha\!=\!0.19\lesssim\alpha_c$ close to the transition.
%\vspace*{-10pt}
}
\label{fig:exp}
\end{figure}

In Fig.~\ref{fig:exp}a we show the $s$-dependence of the exponent $\nu z$.
We find a divergence for both $s\to 0$ and $s\to 1$, consistent
with a Kosterlitz-Thouless transition at $s=1$, and the system being always
localized at $s=0$.
The \mbox{$s\to 1$} divergence is in good agreement with the perturbative
result (\ref{nuzrg}), see below.

The NRG algorithm can be used to compute a variety of static and
dynamic observables.
As an example, we show $C(\omega)$, being the Fourier transform of the
symmetrized autocorrelation function
$C(t) = \frac{1}{2}\langle [\sigma_z(t),\sigma_z]_+ \rangle$,
in Fig.~\ref{fig:exp}b for $s=0.6$
and parameters in the delocalized phase close to the transition.
We observe a crossover from $C(\omega) \propto \omega^s$ at small frequencies,
characteristic of the delocalized phase \cite{KM,spzw}, to
a quantum critical behavior with a power-law divergence at
higher frequencies -- this gives rise to a characteristic peak at $\omega\sim T^\ast$.

%%%%%%%%%%%%%%%%%%%%%%%%%%%%%%%%%%%%%%%%%%%%%%%%%%%%%%%%%%%%%%%%%%%%%%%%%

{\em Comparison to perturbative results. }
%
% As mentioned above,
The partition function of the spin-boson model can be
approximately represented as that of a one-dimensional Ising model with
couplings falling off as $r^{-s-1}$; in this picture, defects in the Ising
system correspond to spin flips of the original spin along the imaginary
time axis.
A RG analysis of this Ising model has been performed by
Kosterlitz \cite{koster}.
Carrying over these results to the spin-boson model, we arrive at the
RG equations (see also Ref.~\onlinecite{dima}):
\begin{eqnarray}
\beta(\alpha) &=& -\alpha ( \bar\Delta^2 + s - 1 ) \,,\nonumber \\
\beta(\bar\Delta) &=& \bar\Delta ( 1 - \alpha ) \,,
\label{eq:rg}
\end{eqnarray}
valid for small $\bar\Delta$,
where $\bar\Delta = \Delta/\omega_c$ is the dimensionless tunneling strength.
The RG flow is sketched in Fig. 1 of Ref.~\onlinecite{koster}.
For $s=1$, these equations are equivalent to the ones known from
the anisotropic Kondo model, and describe a Kosterlitz-Thouless transition,
with a fixed line $\bar\Delta = 0$, $\alpha\geq 1$.
For $s<1$ there is an unstable fixed point at $\alpha=1$, $\bar\Delta^2 = 1-s$;
clearly it is perturbatively accessible for small values of
$(1\!-\!s)$ only.
The critical fixed point is characterized by
\begin{equation}
\nu z = 1/\sqrt{2(1-s)} + {\cal O}(1) \,.
\label{nuzrg}
\end{equation}
For small $\alpha$, $\bar\Delta$, Eq. (\ref{eq:rg}) yields for
the phase boundary
\begin{equation}
\alpha_c \propto \Delta^{1-s} ~~ \mbox{for}~ \Delta\ll\omega_c
\label{alpcrg}
\end{equation}
valid for all $0<s<1$.
The results (\ref{nuzrg}) and (\ref{alpcrg}) are in good agreement with
our numerical data in Figs.~\ref{fig:exp}a and \ref{fig:phd}b, respectively.

%%%%%%%%%%%%%%%%%%%%%%%%%%%%%%%%%%%%%%%%%%%%%%%%%%%%%%%%%%%%%%%%%%%%%%%%%

{\em Conclusions.}
We have presented a generalization of Wilson's NRG to quantum impurity
problems with bosonic baths.
Applying this novel technique to the subohmic spin-boson model,
we have found a line of continuous boundary quantum phase transitions for
all $0<s<1$, with exponents varying as function of $s$.
This line terminates in a Kosterlitz-Thouless transition point at $s=1$.
Near $s=1$, our numerical results are in agreement with perturbative
calculations.
The existence of a transition for $s<1$ implies that weakly damped coherent
dynamics {\em is} possible for qubits coupled to a subohmic bath,
provided that the initial splitting $\Delta$ is large.

In close analogy to the fermionic NRG, our method can be easily applied
to the calculation of dynamical quantities.
Furthermore, generalizations to impurities with multiple bosonic baths
or both fermionic and bosonic baths are possible.
This will allow the study of large classes of impurity models,
e.g., so-called Bose-Kondo and Bose-Fermi-Kondo models.

We thank M. Garst, S. Kehrein, D. Meyer, A. Rosch, and W. Zwerger
for discussions, and in particular W. Zwerger for pointing
out Ref. \onlinecite{koster}.
This research was supported by the DFG through SFB 484 (RB) and
the Center for Functional Nanostructures (MV), and by
the Alexander von Humboldt foundation (NT).

%%%%%%%%%%%%%%%%%%%%%%%%%%%%%%%%%%%%%%%%%%%%%%%%%%%%%%%%%%%%%%%%%%%%%%%%%

%\vspace*{-13pt}


\begin{thebibliography}{}
\bibitem{footnote}
Exceptions are bosonic impurity models allowing a mapping to fermionic ones,
like the ohmic spin-boson model.
\bibitem{states}
In the bosonic NRG the matrices to be diagonalized in each step are of size
$(N_b N_s)^2$, to be compared with $(4 N_s)^2$ for fermionic NRG.
No spin symmetries can be exploited in the spin-boson problem to reduce the matrix size,
in contrast to fermionic Kondo models.
\end{thebibliography}
\end{document}